\documentclass[%
 reprint,
 amsmath,amssymb,
 aps,
]{revtex4-2}

\usepackage{graphicx}
\usepackage{dcolumn}
\usepackage{bm}

\begin{document}

\title{Perturbed Nonlinear Evolution of Optical Soliton Gases: Growth and Decay in Integrable Turbulence}
\author{Loic Fache}
\affiliation{Univ. Lille, CNRS, UMR 8523 - PhLAM -
  Physique des Lasers Atomes et Mol\'ecules, F-59 000 Lille, France}
\author{Fran\c{c}ois Copie}
\affiliation{Univ. Lille, CNRS, UMR 8523 - PhLAM -
  Physique des Lasers Atomes et Mol\'ecules, F-59 000 Lille, France}
\author{Pierre Suret}
\affiliation{Univ. Lille, CNRS, UMR 8523 - PhLAM -
  Physique des Lasers Atomes et Mol\'ecules, F-59 000 Lille, France}
\author{St\'ephane Randoux}
\email{stephane.randoux@univ-lille.fr}
\affiliation{Univ. Lille, CNRS, UMR 8523 - PhLAM -
  Physique des Lasers Atomes et Mol\'ecules, F-59 000 Lille, France}

\date{\today}

\begin{abstract}
We present optical fiber experiments investigating the perturbed, non-integrable evolution of soliton gases (SGs) under weak linear damping and gain. By measuring the amplitude and phase of the optical field in a recirculating loop, we determine the spectral distribution of SGs at various propagation distances. We demonstrate that a SG, initially prepared as a soliton condensate with a Weyl distribution, undergoes significant changes in its spectral distribution due to dissipation. Specifically, weak gain leads to the emergence of a nearly monochromatic SG, while weak damping results in a SG with a semicircular spectral distribution at long propagation distances. These observed dissipation-driven changes in the SG's spectral characteristics are not explained by existing hydrodynamic or kinetic theories.
\end{abstract}


\maketitle

Integrable turbulence (IT) refers to both a physical phenomenon and a theoretical framework for studying the dynamics of nonlinear random waves in systems governed by integrable equations, such as the Korteweg–de Vries (KdV) equation or the one-dimensional nonlinear Schrödinger equation (1D-NLSE) \cite{Zakharov:09,Soto:16,Redor:21,Randoux:14,Michel:20,Congy:24,Randoux:17,Agafontsev:15,Sun:23}. In contrast to wave turbulence \cite{Nazarenko_WT:11,Picozzi:14,Falcon:22,Galtier_book:22}, which addresses the behavior of out-of-equilibrium, nonlinear random dispersive waves in non-integrable systems, the theoretical framework underlying IT is the inverse scattering transform (IST) method \cite{Drazin_book,Ablowitz_book:11}. This distinction provides a relevant perspective on the analysis of large-scale, emergent hydrodynamic phenomena in integrable systems, where soliton gas (SG) serves as a central concept \cite{Zakharov:71,GEl:05,GEL:20,GEL:21,Suret:24}.

SG can be defined as an infinite ensemble of interacting solitons with randomly distributed amplitudes, velocities, and positions \cite{Zakharov:71,GEl:05,Suret:24}. The integrable dynamics, constrained by the existence of an infinite number of conserved quantities, prevent SGs from achieving relaxation to the classical Gibbs ensembles. Instead SGs exhibit local nonthermal stationary states, the so-called generalized Gibbs ensembles, which play a fundamental role in generalized hydrodynamics (GHD), the hydrodynamic theory of many-body quantum and classical integrable systems \cite{Koch:22,Bonnemain:22,Bertini:16,Castro:16,Doyon:25}. Currently, the out-of-equilibrium evolution of SGs represents an active area of research, explored using the spectral kinetic theory of SGs and GHD \cite{GEL:20,Bonnemain:22,Suret:24,Castro:16}.

Despite several experimental observations of SGs across various physical systems \cite{Redor:19,Costa:14,Redor:21,Mossman:24,Marcucci:19,Suret:20,Dieli:24}, only a few experimental studies have questioned the validity of the spectral kinetic theory of SGs in a quantitative way \cite{Fache:24,Suret:23,Fache:24b}. Several experiments have demonstrated that while perturbative higher-order effects are inevitably present, they break integrability to such a negligible degree that the predictions of kinetic theory still align closely with experimental results \cite{Fache:24,Suret:23}. In contrast, other experiments have demonstrated that even small perturbative higher-order effects can significantly alter integrable dynamics, leading to unexpected behaviors not anticipated at the theoretical level \cite{Fache:24b}. For example, a KdV soliton condensate has been shown to emerge spontaneously as a result of weak dissipation in a nonlinear electrical transmission line \cite{Fache:24b}.

In this Letter, we present experiments investigating the perturbed, non-integrable evolution of optical SGs, which reveal unexpected and remarkable changes in their spectral, IST properties. Using a recirculating optical fiber loop that allows precise control of either optical gain or losses, we examine the growth and decay of IT in SG experiments, accurately described by the focusing 1D-NLSE with a weak linear gain or damping term \cite{Agafontsev:20,Agafontsev:23,Coppini:20}. Our experiments reveal that the SG, initially prepared as a soliton condensate characterized by the Weyl distribution, undergoes significant changes in its spectral distribution due to dissipation. We find that a small amount of optical gain leads to the emergence of a nearly monochromatic SG, with most solitons becoming individualized and exhibiting similar spectral (IST) parameters. On the other hand, small damping is found to alter the geometric shape of the IST spectrum, resulting in a semicircular spectral configuration at long evolution time. The dissipation-driven changes observed experimentally in the spectral characteristics of the optical SGs are not explained by existing hydrodynamic or kinetic theories of SG.

Our experimental setup is depicted schematically in Fig. 1. It comprises a recirculating fiber loop constructed from $\sim 8$ km of single-mode fiber (SMF), which is closed upon itself using a $90/10$ fiber coupler. This coupler is configured to recirculate $90\%$ of the optical power. The optical signal circulates in the clockwise direction, and at each round trip, $10\%$ of the circulating power is extracted and directed to a photodetector (PD) connected to a fast oscilloscope with a sampling rate of $160$ GSa/s and a bandwidth of $65$ GHz. The overall detection bandwidth of the oscilloscope and the PD is $32$ GHz. The periodic extraction of light at each round trip within the recirculating fiber loop enables stroboscopic monitoring of the wavefield's evolution every $8$ km. The signals recorded by the oscilloscope are subsequently processed numerically to generate space-time diagrams that illustrate the dynamics of the wavefield over hundreds of round trips inside the fiber loop \cite{Kraych:19a,Kraych:19b,Suret:23}.

\begin{figure}[h]
  \includegraphics[width=8.5cm]{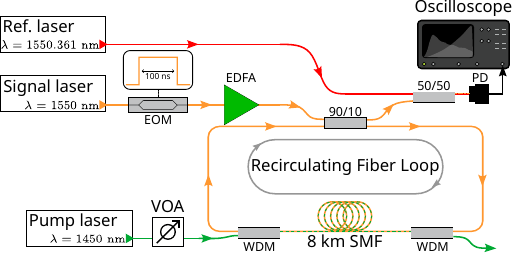}  
  \caption{Experimental setup. A SG, initially in the form of a $100$ ns-long flat-top pulse at $1550$ nm, perturbed by optical noise, propagates through a recirculating fiber loop where the optical gain or loss rate can be precisely controlled through Raman amplification using a $1450$ nm pump laser. Measurement of the amplitude and the phase of the optical signal is made using a heretodyne technique where the optical signal at the output of the fiber loop is beated against a single-frequency laser at $1550.361$ nm. 
}
\end{figure}

A typical experiment involves circulating a long, flat-top pulse ($\sim 100$ ns duration) for $400$ to $800$ round trips in the fiber loop, corresponding to total propagation distances of $\sim 3200$ km to $\sim 6400$ km. The flat-top pulse is generated using a fast electro-optic modulator (EOM) and is perturbed by a small amount of optical noise introduced by an Erbium-doped amplifier prior to the pulse’s injection into the fiber loop, see Fig. 1. As a result of the process of modulation instability, the perturbed flat-top pulse evolves into a fully randomized bound state SG \cite{Gelash:19,Marcucci:19,Bonnefoy:20,GEl:16,Suret:23,Mullyadzhanov:21}. 

The mean gain or loss of the optical signal over hundreds of round trips in the fiber loop can be precisely controlled through Raman amplification, using a counter-propagating 1450 nm pump laser coupled into and out of the loop via wavelength division multiplexers (WDMs), as shown in Fig. 1. By carefully adjusting the optical power of the pump laser using a variable optical attenuator (VOA), we performed experiments where the mean power of the optical SGs is set to exponentially increase or decay at a given rate, allowing us to explore both growing and decaying stages of IT \cite{Agafontsev:20,Agafontsev:23}.

Capturing the changes in the IST spectrum experienced by SGs during propagation is a significant challenge in optics, as it requires simultaneous measurement of both the amplitude and phase of the optical field. In this work, we address this challenge using a heterodyne technique. Specifically, the optical signal at the output of the recirculating fiber loop is mixed with a single-frequency reference field that is detuned by approximately $45$ GHz (equivalently $0.361$ nm) from the central frequency of the optical signal. As illustrated in Fig. 1, the optical signal and the detuned reference field are combined using a $50/50$ fiber coupler, with the output directed to a fast photodetector (PD). The reconstruction of the optical field's amplitude and phase employs conventional signal processing techniques that are described in the Supplemental Material. 

\begin{figure*}[!]
  \includegraphics[width=1\textwidth]{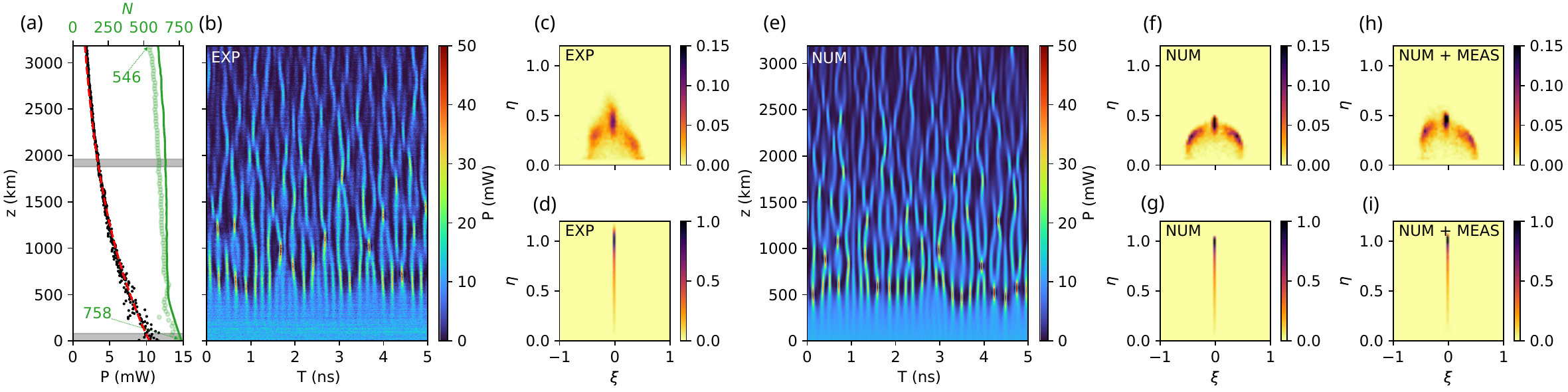}
  \caption{Experiments and numerical simulations showing the evolution of an optical SG in a recirculating fiber loop in the presence of small damping. (a) The black points represent the measured power of the SG as a function of the propagation distance $z$. The green points (resp. line) represent the number of discrete eigenvalues $\lambda_i$ measured in the experiment (resp. computed in numerical simulations) as a function of $z$. (b) Space-time evolution of the optical power of the SG measured in the experiment. (c) DOS of the optical SG measured at $z=1920$ km in the upper complex plane. (d) DOS of the initial condition. (e) Same as (b) but from numerical simulations of Eq. (\ref{eq:NLSE}) with $\gamma=1.3$ W$^{-1}$km$^{-1}$, $\beta_2=-22$ ps$^2$ km$^{-1}$, $\alpha_\text{eff} \sim 5.8 \times 10^{-4}$ km$^{-1}$, $P(z=0)=P_0=11.5$ mW. (f), (g) Same as (c), (d) but computed from simulations reported in (e). (h), (i) Same as (f), (g) but including noise and finite bandwidth effects in the heterodyne measurement, see Supplementary Material for details. 
}
\end{figure*}

Focusing first on experiments involving decaying IT, Fig. 2(b) illustrates the space-time evolution of the optical wavefield power, experimentally observed as the mean power $P(z)$ of the flat-top pulse decreases from $P_0 \sim 11.5$ mW to $\sim 2$ mW over a propagation distance of $\sim 3200$ km (see Fig. 2(a)). At a qualitative level, the experimentally reconstructed space-time diagram over a narrow $5$ ns time window closely resembles the space-time patterns typically associated with the noise-driven modulation instability of plane waves in systems governed by the focusing 1D-NLSE \cite{Toenger:15,Nahri:16,Suret:16,Kimmoun:16,Dudley:19,Kraych:19b,Vanderhaegen:22}.

As shown in Fig. 2(e), the dynamical features observed in our experiment are quantitatively well described by numerical simulations of the following 1D-NLSE with a small linear damping term:
\begin{equation}\label{eq:NLSE}
  i\frac{\partial A}{\partial z}=\frac{\beta_2}{2}\frac{\partial^2
    A}{\partial T^2}-\gamma|A|^2A -i \frac{\alpha_{\rm
      eff}}{2} A.
\end{equation}
$A(z,T)$ represents the complex envelope of the electric field that slowly varies in physical space $z$ and time $T$.  The Kerr coefficient of the fiber is $\gamma=1.3$ W$^{-1}$km$^{-1}$. The group velocity dispersion coefficient is $\beta_2=-22$ ps$^2$ km$^{-1}$. $\alpha_\text{eff}>0$ represents the effective power decay rate of the circulating field. In the experiment presented in Fig. 2(a)(b), its value is $\alpha_\text{eff} \sim 5.8 \times 10^{-4}$ km$^{-1}$ or equivalently $\sim 0.0025$ dB/km. 

Although the space-time diagrams in Figs. 2(b) and 2(e) might suggest that dissipation has a negligible effect on the noise-driven destabilization of the initial flat-top pulse, we now show that the discrete IST spectra of the optical SG reveal distinct features introduced by dissipation to the dynamics. For the computation of the discrete IST spectra, we first introduce the following dimensionless form of the focusing 1D-NLSE  \cite{Coppini:20,Kimmoun:16}
\begin{equation}\label{eq:NLSE_adim}
  i \psi_t+\psi_{xx}+2|\psi|^2 \psi+i\epsilon \psi=0
\end{equation}
which is obtained from Eq. (\ref{eq:NLSE}) using the following transformations: $\psi=A/\sqrt{P_0}$, $x=T\sqrt{\gamma P_0/|\beta_2|}$, $t= \gamma P_0 z /2$, $\epsilon=\alpha_\text{eff}/(\gamma P_0)$ \cite{Suret:23}.

Within the IST formalism, the non-self-adjoint Zakharov-Shabat eigenvalue problem that is associated with Eq. (\ref{eq:NLSE_adim}) for $\epsilon=0$ reads \cite{yang2010nonlinear}
\begin{equation}\label{ZS}
\widehat{\bf \mathcal{L}} \Phi = \lambda \Phi , \qquad
\widehat{\bf \mathcal{L}} =
 \begin{pmatrix}
 - i \partial_x  & -i \psi \\ - i \psi^* & i \partial_x \\ 
 \end{pmatrix}
 .
\end{equation}
${\bf \Phi}(x, \lambda)$ is a vector wave function. $\lambda \in \mathbb{C}$ represent the eigenvalues composing the discrete spectrum associated with the soliton content of the field $\psi=\psi(x,t)$ that is measured at some given evolution time $t$ (equivalently at some given propagation distance $z$ in the experiment). Each soliton in a soliton gas, consisting of an ensemble of $N$ solitons, is characterized by a discrete eigenvalue $\lambda_j=\xi_j+i\eta_j \, (j=1,..,N)$ of the spectrum of the linear operator $\bf \widehat{\mathcal{L}}$, where $\xi_j$ and $\eta_j$ parameterize the velocity and amplitude of each soliton, respectively. Importantly, integrable dynamical evolution (implying $\epsilon=0$ in Eq. (\ref{eq:NLSE_adim})) entails an isospectral property, meaning that all the discrete eigenvalues $\lambda_j$ remain invariant with respect to the evolution variable $t$.

The discrete IST spectra of the optical SG are computed from the normalized experimental complex fields $\psi=A/\sqrt{P_0}$ by solving numerically Eq. (\ref{ZS}) using the Fourier collocation method described in ref. \cite{yang2010nonlinear}. The output of this numerical calculation is an ensemble of several hundreds of discrete, complex-valued eigenvalues $\lambda_j$ that are located in a bounded region of the upper complex plane. Given the large number of discrete eigenvalues that are found from nonlinear spectral analysis, we follow the approach used in the kinetic theory of SG and compute the so-called density of states (DOS) $f(\lambda; x,t)$ of the SG.

The DOS $f(\lambda; x,t)$, where $\lambda=\xi + i \eta$, represents the density of soliton states in the phase space, i.e. $f d\xi d\eta d x$ is the number of solitons contained in a portion of  SG with the complex spectral parameter $\lambda \in [\xi, \xi+d\xi] \times [\eta, \eta +d\eta]$ over the space interval $[x, x+dx]$  at time $t$ (corresponding to the position $z$ in the fiber loop experiment). Given that the optical SG is spatially homogeneous, the DOS represents the probability density function of the complex-valued discrete eigenvalues normalised in such a way that $ \int_{-\infty}^{+\infty} d\xi \int_{0}^{+\infty} d \eta \, f(\lambda)  =N/\Delta x$, where $N$ represents the number of eigenvalues found in the upper complex plane and $\Delta x$ represents the spatial extent of the gas \cite{Suret:20,GEL:21}.

In practice, the experimental DOS is computed not from a single flat-top pulse, but from three identical flat-top pulses, each evolving independently within the fiber loop during the same experimental run. Assuming minimal variation in the IST spectra over short propagation distances, the DOS is calculated by averaging the IST spectra of the three pulses collected over a $50$ km propagation span, as highlighted by the grey-shaded regions in Fig. 2(a).

Fig. 2(d) shows the DOS of the flat-top pulse perturbed by small optical noise that is taken as initial condition in the experiment ($z=0$ km). As it can be easily anticipated from previous works \cite{Gelash:19}, we find that the measured DOS is very well fitted by the Weyl distribution $f_W(\eta)=\eta/(\pi\sqrt{1-\eta^2})$ located along the vertical imaginary axis $\xi=0$ (see also Fig. 4(a)). Due to dissipation, the isospectrality condition characteristic of integrable dynamics is not expected to be satisfied in our experiment. However, since the damping effects are small ($\epsilon \simeq 0.038$ in Eq. (\ref{eq:NLSE_adim})), IST analysis remains a relevant tool for examining the evolution of the optical SG. Fig. 2(c), computed at $z \sim 1900$ km, shows that optical damping transforms the spectral support of the DOS in the upper complex plane, reshaping the vertical linear support of the initial Weyl distribution into a semicircular one. It is important to note that the concept of a circular spectral support for the DOS of a SG was introduced mathematically in Ref. \cite{GEL:20}, but without any expectation that this class of SGs could be observed experimentally nor that it could emerge from a non-integrable evolution.

\begin{figure*}[!]
  \includegraphics[width=1\textwidth]{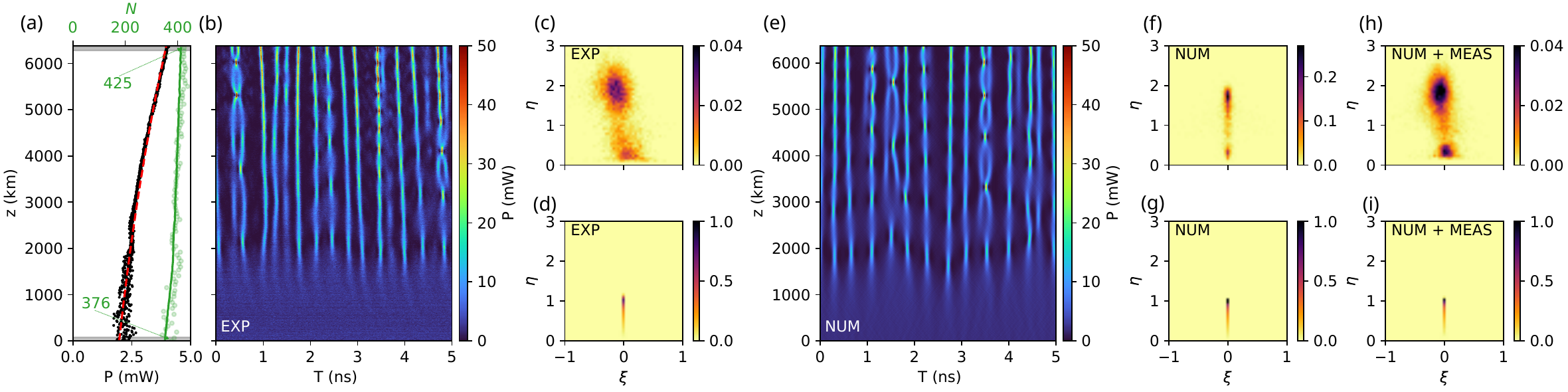}
  \caption{Experiments and numerical simulations showing the evolution of an optical SG in a recirculating fiber loop in the presence of small gain. (a) The black points represent the measured power of the SG as a function of the propagation distance $z$. The green points (resp. line) represent the number of discrete eigenvalues $\lambda_i$ measured in the experiment (resp. computed in numerical simulations) as a function of $z$. (b) Space-time evolution of the optical power of the SG measured in the experiment. (c) DOS of the optical SG measured at $z=6320$ km in the upper complex plane. (d) DOS of the initial condition. (e) Same as (b) but from numerical simulations of Eq. (\ref{eq:NLSE}) with $\alpha_\text{eff} \sim - 1.1 \times 10^{-4}$ km$^{-1}$, $P(z=0)=P_0=2$ mW. (f), (g) Same as (c), (d) but computed from simulations reported in (e). (h), (i) Same as (f), (g) but including noise and finite bandwidth effects in the heterodyne measurement, see Supplementary Material for details.
}
\end{figure*}

As shown in Fig. 2(f) and 2(g), the dissipation-driven changes of the geometrical support of the DOS are also found in numerical simulations of Eq. (\ref{eq:NLSE}) that are made with the experimental parameters. It is noteworthy that the DOS obtained from numerical simulations of Eq. (\ref{eq:NLSE}), shown in Fig. 2(f), exhibits a perfectly symmetric shape that is not found in the experiments, as seen in Fig. 2(c). This slight discrepancy between the experimental and numerical DOS, along with the blurrier appearance of the experimental DOS, is attributed to noise and finite bandwidth effects in the heterodyne measurement of the optical field's phase and amplitude, as illustrated in Fig. 2(h) and 2(i) and discussed more in details in Supplementary Material.

Next, we examine another dissipative experiment in which the optical SG undergoes a linearly amplified evolution instead of a damped one. Increasing the power of the $1450$ nm pump laser, we now consider a situation in which the mean optical power of the SG increases from $\sim 2$ mW to $\sim 4$ mW over $\sim 6300$ km, as shown in Fig. 3(a). Fig. 3(b) reveals that the space-time evolution of the optical SG looks qualitatively different from the situation with damping reported in Fig. 2(b). Following the initial destabilization of the flat-top pulse, individual nonlinear structures with small relative velocities are observed to form and persist within the fiber loop. At a qualitative level, the observed space-time evolution shows remarkable similarities to that reported in previous numerical studies on growing IT, even though these studies considered a completely different initial condition, starting from a Gaussian noise of small amplitude \cite{Agafontsev:23}.

Fig. 3(c) and 3(d) show that the DOS of the SG, initially in the form of the Weyl distribution, essentially becomes a clustered distribution centered around $\lambda_0 \simeq 1.8i$. The fact that $Re(\lambda_0) \simeq 0$ is the spectral signature of all solitons in the amplified SG exhibiting small velocities, in stark contrast to the scenario depicted in Fig. 2(c), where damping was shown to broaden the soliton velocity distribution. The fact that $Im(\lambda_0) \simeq 1.8$ means that most of the solitons in the SG formed at $z \sim 6300$ km have nearly identical amplitudes, mirroring a nearly monochromatic SG \cite{GEl:05,Fache:24}.

Fig. 3(e) shows that the space time evolution observed in the experiment is well reproduced by numerical simulations of Eq. (\ref{eq:NLSE}). Fig. 3(f)-(i) show that the features observed experimentally in terms of evolution of the DOS are also found in numerical simulations that include noise and finite bandwidth effects due to the heterodyne measurement of the optical field's phase and amplitude, see Supplementary Material for details. 

In our experiments, dissipation significantly alters the mass $M(t)=\int_{-\infty}^{+\infty} |\psi(x,t)|^2 dx$, or equivalently, the mean power, $P(z)=\int_{-\infty}^{+\infty} |A(T,z)|^2 dT$, of the optical SG, see Fig. 2(a) and 3(a). However, the total number of discrete eigenvalues, $N(z)$, does not vary in proportion to the changes in mass. In Fig. 2(a) the mass decreases by a factor $\sim 6$, while $N$ changes by only $\sim 20 \%$ during the evolution. Similarly, in Fig. 3(a), the mass approximately doubles but $N$ changes by only $10 \%$ during the evolution. This indicates that the changes in the mass of the optical SGs are not primarily caused by the creation or annihilation of soliton states, but rather by the rearrangement of the DOS in the upper complex plane.

To substantiate this point in a more quantitative way, we consider the expression of the mass of SG given by the spectral kinetic theory: 
\begin{equation}\label{eq:mass}
  M = 4 \int_{0}^{+\infty}  \eta \, \tilde{f}(\eta) \, d\eta   =  \int_{0}^{+\infty}  \rho(\eta) \, d\eta. 
\end{equation}
where $\tilde{f}(\eta) = \int_{-\infty}^{+\infty} d\xi \,  f(\xi,\eta)$ represents the DOS $f(\xi,\eta)$ that has been integrated over the real axis $\xi$ \cite{GEL:20,GEL:21}. $\rho(\eta)= 4 \eta \tilde{f}(\eta)$ represents the spectral density of mass with respect to $\eta$. 

\begin{figure}[!]
  \includegraphics[width=0.45\textwidth]{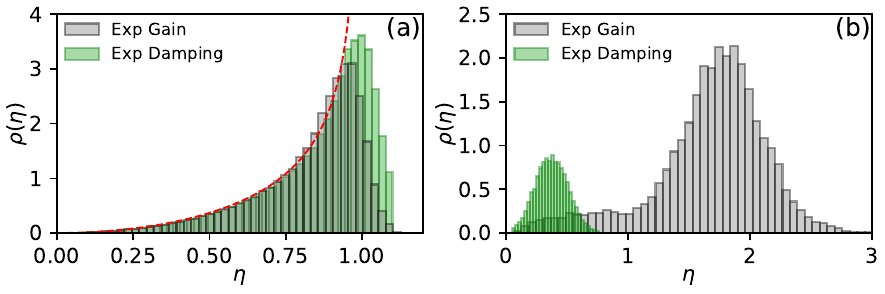}
  \caption{Experimental histograms showing the spectral density of mass, $\rho(\eta)$, along the vertical imaginary axis $\eta$. (a) The green (resp. grey) histogram represents the initial distribution in the experiment with linear damping (resp. amplification) of the SG. The red dashed line represents the spectral density of mass $\rho_W(\eta)=4\eta f_W(\eta)$ associated with the Weyl distribution $f_W(\eta)=\eta/(\pi\sqrt{1-\eta^2})$. (b) Same as (a) but at $z \sim 1920$ km (green histogram) and $z \sim 6320$ km (grey histogram). 
}
\end{figure}

Fig. 4(a) illustrates that the spectral density of mass, $\rho(\eta)$, for the initial bound-state SG closely matches the spectral density $\rho_W(\eta)=4\eta f_W(\eta)$, which corresponds to the Weyl distribution. In contrast, Fig. 4(b) shows how experimental damping significantly alters $\rho(\eta)$, transforming it into a bell-shaped distribution centered around $\eta \simeq 0.4$ (green histogram). On the other hand, linear amplification of the SGs results in the development of an asymmetric distribution of the spectral density (grey histogram in Fig. 4(b)), with its peak shifted to approximately $\eta \simeq 1.8$.

In conclusion, we have presented optical fiber experiments exploring the perturbed evolution of SGs under the influence of small linear damping and gain. Our experiments reveal that dissipation-induced changes in the mass (equivalently, in the mean power) of optical SGs are predominantly driven by a redistribution of the DOS in the upper complex plane, rather than by the creation or annihilation of soliton states. Previous studies investigating dissipative evolution in NLS systems have primarily focused on ensembles containing only a few solitons, employing approaches such as numerical simulations or perturbed IST theory \cite{Blow:88,Okamawari:95,Prilepsky:07,Bohm:07,Hause:18}. We hope that our experiments, which explore large ensembles of interacting solitons with random characteristics, will inspire theoretical investigations using GHD, possibly building on previous studies of one-dimensional Bose gases with atom losses \cite{Bouchoule:20,Bouchoule:21}. Furthermore, our experiments demonstrate that higher-order perturbative effects can be exploited to manipulate the spectral (IST) characteristics of SGs, offering new experimental opportunities.

\begin{acknowledgments}
  We thank Dmitry Agafontsev, Thibault Bonnemain, Thibault Congy, Gennady El, Andrey Gelash, Giacomo Roberti, and Alexander Tovbis for their collaboration and the discussions we have had over the years on SG-related topics. This work has been partially supported  by the Agence Nationale de la Recherche  through the SOGOOD (ANR-21-CE30-0061) projects, the LABEX CEMPI project (ANR-11-LABX-0007), the Ministry of Higher Education and Research, Hauts de France council and European Regional Development Fund (ERDF) through the Nord-Pas de Calais Regional Research Council and the European Regional Development Fund (ERDF) through the Contrat de Projets Etat-R\'egion (CPER Photonics for Society P4S). The authors would like to thank the Centre d'Etudes et de Recherche Lasers et Application (CERLA) for technical support and the Isaac Newton Institute for Mathematical Sciences, Cambridge, for support and hospitality during the programme Emergent phenomena in nonlinear dispersive waves, where work on this paper was undertaken. This work was supported by EPSRC grant EP/V521929/1.
\end{acknowledgments}





\bibliographystyle{apsrev4-1}
%

\pagebreak
\newpage

\renewcommand{\theequation}{S\arabic{equation}}
\setcounter{equation}{0}
\onecolumngrid


\renewcommand{\theequation}{S\arabic{equation}}
\renewcommand{\thefigure}{S\arabic{figure}}

\vspace{14cm}

\begin{center}
  {\bf Supplemental material for : \\"Perturbed Nonlinear Evolution of Optical Soliton Gases: Growth and Decay in
Integrable Turbulence"}\\
\end{center}

\begin{center}
    Loic Fache,$^1$ Fran\c{c}ois Copie,$^1$ Pierre Suret,$^1$ and  St\'ephane Randoux$^1$
\end{center}

\begin{center}
  {\it $^1$ Univ. Lille, CNRS, UMR 8523 - PhLAM -Physique des Lasers Atomes et Mol\'ecules, F-59 000 Lille, France}
\end{center}

The purpose of this Supplemental Material is to provide some details about the experimental setup and about the experimental methodology.  All equations, figures, and reference numbers within this document are prepended with ``S'' to distinguish them from corresponding numbers in the Letter.\\
\tableofcontents
\section{\label{sec0} Detailed description of the experimental setup}\label{sec:setup}

\begin{figure}[h!]
    \centering
    \includegraphics[width=\textwidth]{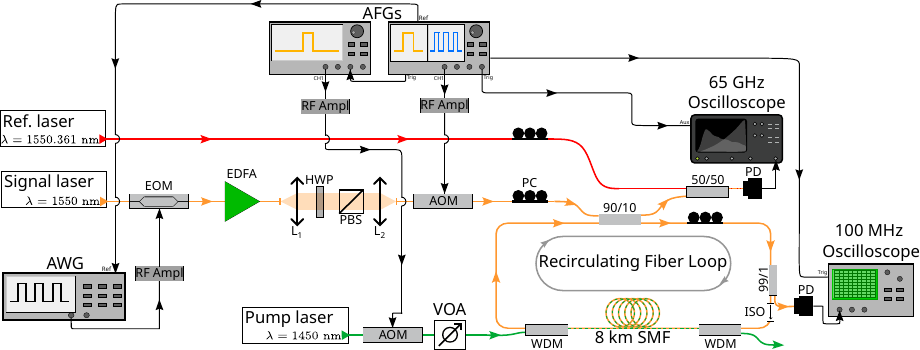}
    \caption{Detailed representation of the experimental setup. AFG: Arbitrary function generator. AWG: Arbitrary waveform generator. EDFA: Erbium-doped fiber amplifier. PBS: Polarizing beam splitter. HWP: Half-wave plate. AOM: Acousto-optic modulator. EOM: Electro-optic modulator. VOA: Variable optical attenuator. }
    \label{fig0}
\end{figure}

Fig.~\ref{fig0} shows a detailed representation of our experimental setup, including some technical details about the generation of the optical signals that are not described in the Letter. The light source used for generation of optical soliton gases (SGs) is a single-frequency continuous-wave (CW) laser diode (APEX-AP3350A) centered at $\lambda$ = 1550 nm which delivers an optical power of a few milliwatts. The optical SGs have the initial form of flat-top pulses with durations of 100 ns. These pulses are produced by using a 20-GHz electro-optic modulator (EOM, iXblue MX-LN-20) connected to an arbitrary waveform generator (AWG, Tektronix AWG70001A) having a bandwidth of 15 GHz. \\

The flat-top pulses are amplified at the Watt-level by using a Erbium-doped fiber amplifier (EDFA, Keopsys CEFA-C-BO-HP). Importantly the amplified spontaneous emission (ASE) from the erbium-doped fiber amplifier (EDFA) introduces a small amount of optical noise to the flat-top pulses. This noise drives their transformation into fully randomized soliton gases as they propagate nonlinearly within the recirculating fiber loop.\\

The mean power of the light signal at the output of the EDFA is regulated using bulk optics, which include a polarization beam splitter (PBS) and a half-wave plate (HWP). Following this power adjustment, the light signal is directed to an acousto-optic modulator (AOM, VSF MT110-IR25-Fio). The AOM functions as an optical gate, remaining open for approximately 500 ns to allow the injection of the light signal into the fiber loop. It then remains closed for significantly longer periods, typically around 5 ms, during which the signal circulates within the fiber loop. The gating signals for the AOM are generated by a 250-MHz arbitrary function generator (AFG, Tektronix 31252). \\

The recirculating fiber loop comprises approximately $8$ km of single-mode fiber (SMF), configured as a closed loop using a $90/10$ fiber coupler. The SMF, manufactured by Draka-Prysmian, has a measured second-order dispersion coefficient of $\beta_2$ = -22 ps$^2$ km$^{-1}$ and an estimated Kerr coefficient $\gamma$ = 1.3 km$^{-1}$W$^{-1}$ at the working wavelength of $1550$ nm. The 90/10 coupler is arranged so that $90\%$ of the intra-loop optical power is recirculated. At each round trip, $10\%$ of the circulating power is extracted and directed to a $50/50$ fiber coupler, where it is mixed with a detuned CW reference laser for heterodyne measurement. The output of the fiber coupler is then routed to a photodetector (Finisar XPDV2120R), which is coupled to a sampling oscilloscope (Teledyne LeCroy LabMaster 10-65Zi, 160 GSa/s). This setup provides an overall detection bandwidth of $32$ GHz.\\

The experimental data comprise a series of temporal sequences, each representing $500$ ns per round trip, triggered via the second channel of the AFG. These sequences are processed numerically to construct single-shot space-time diagrams that reveal the wavefield dynamics. The evolution of the mean power within the loop is carefully controlled through Raman amplification via a counter-propagating pump laser at $1450$ nm. This $1450$ nm laser is a commercial Raman fiber laser (IPG FiberTech) capable of delivering several watts of optical power. It is coupled into and out of the loop through wavelength division multiplexers (WDMs). For the experiments, the power at $1450$ nm is attenuated to approximately $200$ mW using a $90/10$ fiber coupler (not shown in Fig.~\ref{fig0}). The fine control of the pump power injected within the recirculating fiber loop is made using an additional AOM together with a variable optical attenuator (VOA). Note that a slow photodetector (Thorlabs DET10C/M) connected to a $100$-MHz oscilloscope (Agilent 54624A) is used to monitor the slow (ms scale) evolution of the power of the soliton gas within the fiber loop. \\

The heterodyne measurement of the optical signal’s amplitude and phase is performed by mixing it with a reference signal generated by a single-frequency CW laser diode (APEX-AP3350A) operating at $\lambda + \delta \lambda= 1550.361$ nm, as detailed further in Sec. \ref{sec1}.\\

\section{\label{sec1} Heterodyne measurement of the amplitude and phase of the optical field }\label{sec:heterodyne}

\begin{figure}[!h]
    \centering
    \includegraphics[width=\textwidth]{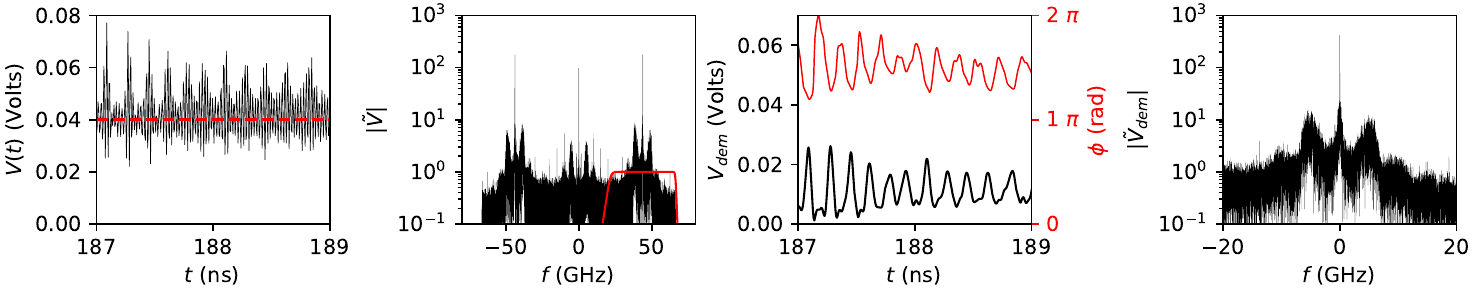}
    \caption{Demodulation scheme of the detected optical signal. (a) Typical experimental signal measured by the photodiode and the oscilloscope at $z=400$ km over $2$ ns. The red dashed line represents the mean value of the detected signal. (b) Fourier power spectrum of the signal plotted in (a).  The red line represents the modulus of the transfer function of the passband Butterworth filter used to filter the frequency components centered around $45$ GHz. (c) Black line : amplitude of the demodulated signal $A_{sig}(t)$. Red line: phase $\phi(t)$ of the optical signal. (d) Fourier power spectrum associated to the demodulated signal $A_{sig}(t) exp(i \phi(t))$ shown in (c). }
    \label{fig2}
\end{figure}

As shown in Fig. S1, the fast PD  placed at the output of the recirculating fiber loop detects the beating between the optical signal at $1550$ nm and a frequency-detuned reference laser at a wavelength of $1550.361$ nm (see Fig.~\ref{fig2}(a)). The reference laser (APEX-AP3350A) is a single-frequency laser diode with an output power of a few mW and an intrinsic linewidth of 300 kHz (constructor datasheet). The wavelength difference ($0.361$ nm) between the two lasers corresponds to a frequency difference $\Delta f_0 \sim 45$ GHz at which the beating between the signal laser and the reference laser is detected by the fast PD (see Fig.~\ref{fig2}(b)). The voltage detected by the PD is given by:
\begin{equation}\label{eq:p}
    V(t) = k  \left| A_{\text{sig}}(t) e^{i\phi(t)} + A_{\text{Ref}} e^{i (2\pi \Delta f_0 t + \phi_{\text{Ref}})} \right|^2
\end{equation}
where $k$ is a real constant whose value is associated with the efficiency of the PD.  $A_{\text{sig}}(t)$ (resp. $\phi(t)$) represents the slowly-varying amplitude (resp. the phase) of the signal field. $A_{\text{Ref}}$ (resp. $\phi_{\text{Ref}}$) represents the amplitude (resp. the phase) of the reference field. Both $A_{\text{Ref}}$ and $\phi_{\text{Ref}}$ are assumed to remain constant during the measurement period, which spans 100 ns—the duration of the flat-top pulses corresponding to the extent of the optical soliton gas in the experiment. \\

Expanding Eq. (\ref{eq:p}), we get:
\begin{equation}
  V(t) = k \left(  A_{\text{sig}}^2(t)  +  A_{\text{Ref}}^2  + A_{\text{sig}}(t) A_{\text{Ref}} e^{i (2\pi \Delta f_0 t + \phi_{\text{Ref}} - \phi(t))} + A_{\text{sig}}(t) A_{\text{Ref}} e^{-i (2 \pi \Delta f_0 t + \phi_{\text{Ref}} - \phi(t))} \right)
\end{equation}

This expression shows that the Fourier spectrum of the detected voltage is composed of three components: one DC component and two side-bands at $\pm \Delta f_0 \sim \pm 45$ GHz, as shown in Fig.~\ref{fig2}(b). Considering that $\Delta f_0$ is sufficiently large that the overlap between the $3$ spectral components can be neglected, we use a highly selective frequency filter to eliminate the DC component and the frequency components around $-45$ GHz. The filtering operation is purely numerical; it is applied at the post-processing stage on the signal recorded by the fast oscilloscope. We have used a causal 15th-order Butterworth filter with a bandwidth of \(\Omega = 40 \text{ GHz}\) (full-width at half maximum) centered at \(\Delta f_0 = + 45 \text{ GHz}\). The modulus of the transfer function of this filter is plotted in red line in Fig.~\ref{fig2}(b).\\

The electrical signal $V_F(t)$ at the output of the filter reads
\begin{equation}
  V_F(t) = k\left(  A_{\text{sig}}(t) A_{\text{Ref}} e^{i (2 \pi \Delta f_0 t + \phi_{\text{Ref}} - \phi(t))} \right)
\end{equation}
The next step in the signal processing is a demodulation operation in which the filtered signal $V_F(t)$ is multiplied by \(e^{-2i \pi \Delta f_0 t}\), which results into a demodulated signal $V_{\text{dem}}(t)$ that reads :
\begin{equation}\label{eq:pdem}
  V_{\text{dem}}(t) = k \left(  A_{\text{sig}}(t) A_{\text{Ref}} e^{i ( \phi_{\text{Ref}} - \phi(t))} \right)
\end{equation}

Removing the constant phase $\phi_{\text{Ref}}$ of the reference laser, the phase of the optical signal is extracted numerically by taking the argument of $V_{\text{dem}}(t)$ in Eq. (\ref{eq:pdem}):
\begin{equation}\label{eq:phase_sig}
  \phi(t) = -\text{Arg} (V_{\text{dem}}(t))
\end{equation}

The last step in the measurement procedure  consists in converting the voltage $V_{\text{dem}}(t)$ into the power $A_{\text{sig}}^2(t)$ of the signal field. Following the experimental approach already used in ref. \cite{Suret:23}, we take advantage of the well-known process of noise-induced modulation instability (MI) of plane waves in the focusing propagation regime \cite{Toenger:15,Nahri:16,Suret:16,Dudley:19,Kraych:19b,Vanderhaegen:22}. After some propagation distance within the fiber loop, the flat-top square pulses perturbed by the optical noise of the EDFA are destabilized, which results in the emergence of a random train of coherent structures. In the Fourier spectrum, this phenomenon appears as the growth of two spectral sidebands symmetrically shifted by $\pm \Omega_\text{max}$ relative to the carrier frequency of the plane wave (see Fig.~\ref{fig2}(d)). The frequency $\Omega_\text{max}$ is related to the power $P_0$ of the plane wave (or of the flat-top pulse) through the relation $\Omega_\text{max}=(2 \gamma P_0/|\beta_2|)^{1/2}$. 

In practice, measuring $\Omega_\text{max}$ is more precise when performed in the time domain by simply counting the peaks of the structures that emerge from modulation instability (MI) over a long observation window of several tens of nanoseconds. Using this approach, we determine the period of MI as $T_\text{MI} = 2 \pi /\Omega_\text{max}=190.1$ ps. This corresponds to an optical power of $P_0 = (2 \pi/T_\text{MI})^2 |\beta_2 |/(2 \gamma) = 9.24$ mW. However, this measurement, taken at $z=400$ km (after the full development of MI), needs to be corrected to account for the optical losses in the fiber loop. Since the power in the loop decays exponentially, the corrected value is $P_0 \simeq 11.5$ mW. This calibrated value is crucial, as it enables the conversion of the voltage measured by the fast PD into an equivalent optical power.

\section{\label{sec2} Influence of finite bandwidth effects and of detection noise on the measurement of the DOS of the optical soliton gas} 


In this section, we employ numerical simulations to demonstrate how noise and finite bandwidth effects, intrinsic to our experimental detection system, introduce distortions in the measured amplitude and phase of the optical signal. These distortions, in turn, impact the measured IST spectrum and the density of states (DOS) of the optical soliton gas.\\ 


\textbf{Influence of the heterodyne detection scheme} \\

\begin{figure}[!h]
    \centering
    \includegraphics[width=\textwidth]{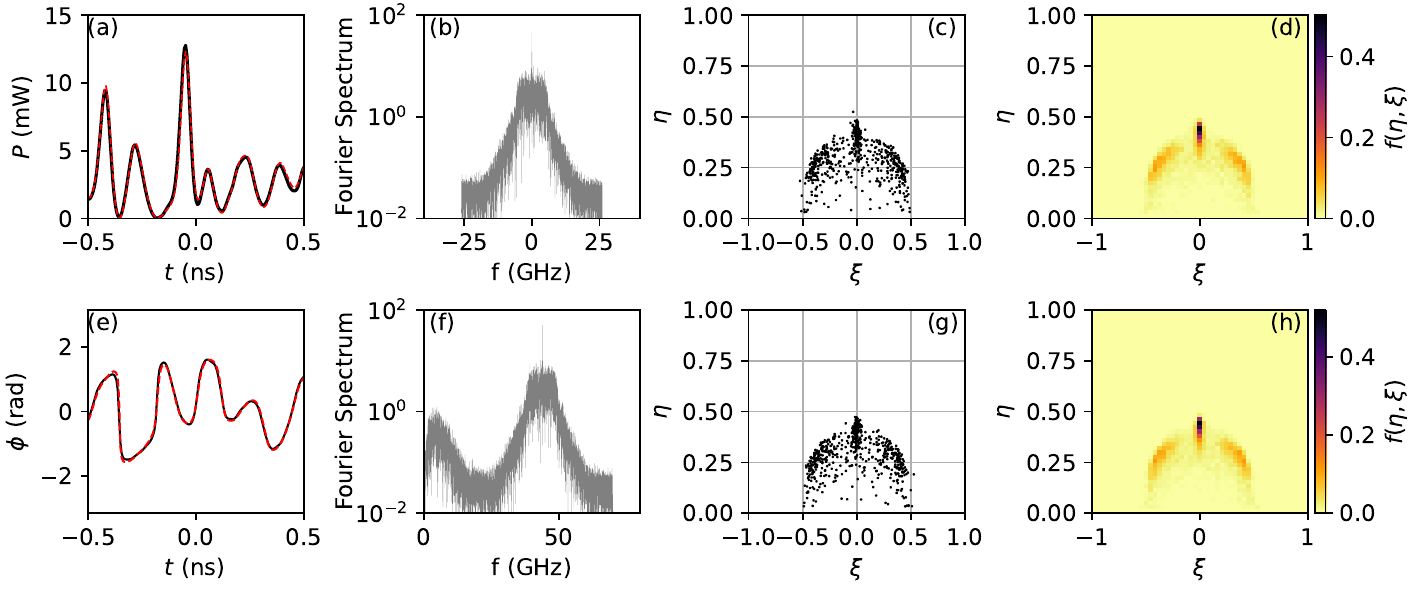}
    \caption{(a) The black line represents the power fluctuations of the optical SG obtained from numerical simulation of NLSE with damping at $z=1920$ km (Eq. (1) of the Letter). The red dashed line represents power fluctuations that are reconstructed using the heterodyne scheme described in Sec. \ref{sec:heterodyne}. (b) Fourier power spectrum of the optical SG. (c) IST spectrum of the optical SG. (d) DOS computed from 10 realizations of the SG. (e) Same as in (a) but for the optical phase of the SG. (f) Fourier power spectrum of the heterodyne signal at $45$ GHz. (g) IST spectrum of the optical SG reconstructed from heterodyne measurement. (h) DOS of the optical SG reconstructed from the heterodyne measurement. }
    \label{fig3_0}
\end{figure}

As a preliminary step, we first demonstrate, using Fig.~\ref{fig3_0}, that the heterodyne detection scheme described in Sec. \ref{sec:heterodyne} preserves the accuracy of the measurements of the IST spectrum and the density of states (DOS) of the soliton gas. \\

Fig.~\ref{fig3_0}(a) represents numerical simulations of the damped NLSE (Eq. (1) of the Letter) showing typical power fluctuations of the optical SG at $z=1920$ km. Fig.~\ref{fig3_0}(b) and \ref{fig3_0}(c) represent the associated Fourier and IST spectra. Fig.~\ref{fig3_0}(d) represents the DOS that has been calculated using 10 numerical realizations of the optical SG. \\

Fig.~\ref{fig3_0}(g) and \ref{fig3_0}(h) in the bottom row of Fig. \ref{fig3_0} demonstrate that the IST spectrum and the DOS remain unaffected by the heterodyne detection scheme described in Sec. \ref{sec:heterodyne}, which essentially involves shifting the Fourier spectrum around $45$ GHz, as illustrated in Fig. \ref{fig3_0}(f). This invariance is also clearly evident in Fig.~\ref{fig3_0}(a) and \ref{fig3_0}(e), where the red and black lines show that the amplitude and phase reconstructed from the heterodyne scheme closely match the numerical counterparts.\\

\textbf{Influence of the finite detection bandwidth} \\

\begin{figure}[h!]
    \centering
    \includegraphics[width=\textwidth]{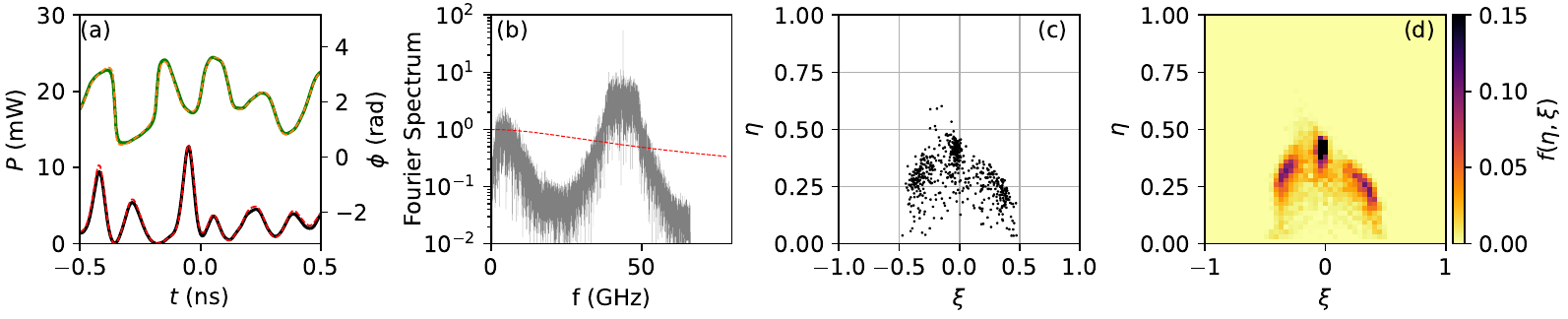}
    \caption{(a) The black (resp. blue) line represents the power (resp. phase) fluctuations of the optical SG obtained from numerical simulation of NLSE with damping at $z=1920$ km (Eq. (1) of the Letter). The red (resp. orange) dashed line represents power (resp. phase) fluctuations that are reconstructed using the heterodyne scheme including finite bandwidth effects. (b) Fourier power spectrum of the heterodyne signal. The red dashed line represents the modulus of the transfer funtion of the $32$ GHz filter describing finite bandwidth effects. (g) IST spectrum of the optical SG reconstructed from heterodyne measurement with finite bandwith effects. (h) DOS of the optical SG reconstructed from the heterodyne measurement with finite bandwith effects. }
    \label{fig3_1}
\end{figure}

We now demonstrate that the finite detection bandwidth of our experimental system introduces a slight distortion in the IST spectrum, which in turn affects the measured DOS of the optical soliton gas. \\

As described in Sec. \ref{sec:setup}, the power fluctuations of the SG are recorded using a fast photodiode (Finisar XPDV2120R) with a nominal bandwidth of 50 GHz, connected to a fast oscilloscope (LeCroy Labmaster 10-65ZI) with a nominal bandwidth of 65 GHz and a sampling rate of 160 GSa/s. Together, these devices behave as a first-order electrical filter having a transfer function given by 
\begin{equation}
    H(f) = \frac{1}{1 + i\frac{f}{f_c}},
    \label{transfer_func}
\end{equation}
with $f_c = 32 \, \text{GHz}$.  \\

The modulus of this transfer function is represented by the red dashed line in Fig.~\ref{fig3_1}(b), illustrating the effect of the low-pass filter on the frequency-shifted signal at $45$ GHz. As shown in Fig.~\ref{fig3_1}(a), the power and phase of the signal reconstructed from the filtered heterodyne signal closely match the numerical profiles that would result without any finite bandwidth effects. While these differences due to the finite bandwidth appear minimal, the computed IST spectrum in Fig.~\ref{fig3_1}(c) reveals an asymmetry, which becomes even clearer in the DOS of the optical SG, as depicted in Fig.\ref{fig3_1}(d).\\

\textbf{Influence of detection  noise}\\

\begin{figure}[h!]
    \centering
    \includegraphics[width=\textwidth]{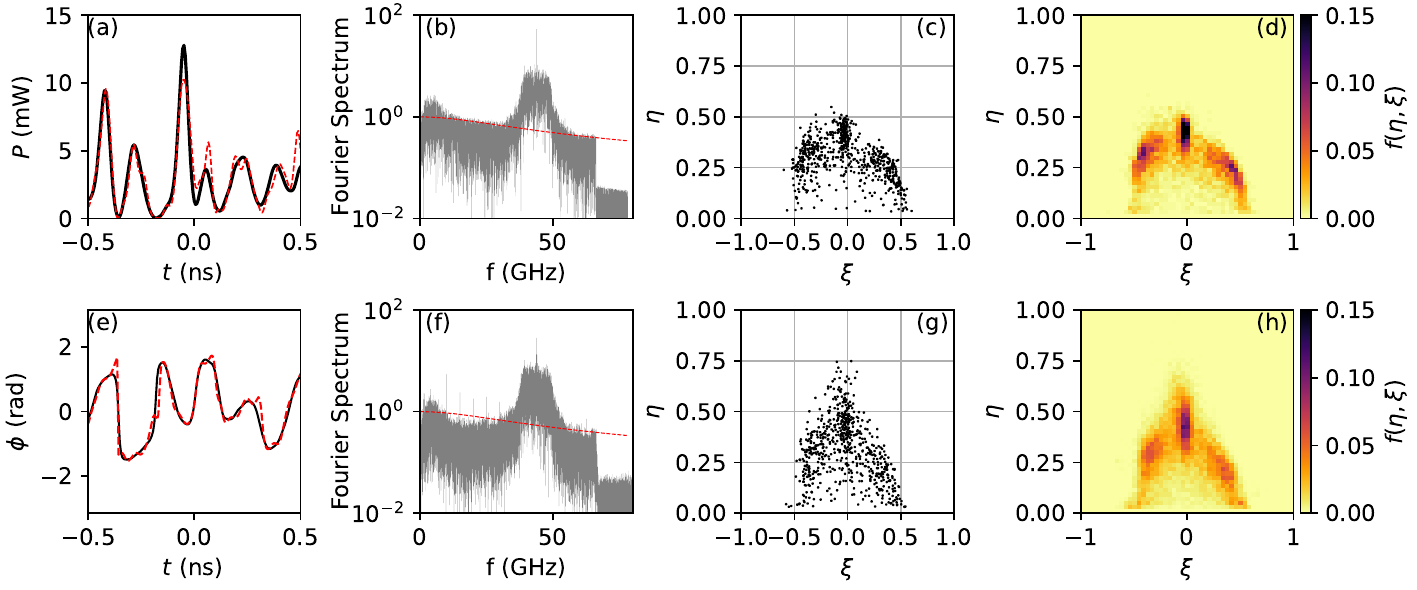}
    \caption{(a) The black line represents the power fluctuations of the optical SG obtained from numerical simulation of NLSE with damping at $z=1920$ km (Eq. (1) of the Letter). The red  dashed line represents power fluctuations that are reconstructed using the heterodyne scheme including finite bandwidth effects and detection noise. (b) Fourier power spectrum of the heterodyne signal perturbated by the added detection noise. The red dashed line represents the modulus of the transfer function of the $32$ GHz filter describing finite bandwidth effects. (c) IST spectrum of the optical SG reconstructed from heterodyne measurement with noise and finite bandwith effects. (d) DOS of the optical SG reconstructed from the heterodyne measurement with noise and finite bandwith effects. (e) The black line represents the phase fluctuations of the optical SG obtained from numerical simulation of NLSE with damping at $z=1920$ km. The red dashed line represents phase fluctuations that are reconstructed using the heterodyne scheme including finite bandwidth effects and detection noise. (f) Fourier power spectrum of the heterodyne signal recorded in the experiment at $z=1920$ km. (g) Associated IST spectrum of the experimental field. (h) DOS experimentally measured at $z=1920$ km.}
    \label{fig3_3}
\end{figure}

In the last step, we consider the infuence of the detection noise of the PD on the measured IST spectra and DOS. \\

Fig.~\ref{fig3_3}(b) shows that we have modified the Fourier spectrum of the heterodyne signal by adding some noise having a level in the Fourier domain that is similar to the experimental detection noise, compare Fig.~\ref{fig3_3}(b) with Fig.~\ref{fig3_3}(f) and also with Fig.~\ref{fig3_1}(b). As a result of the incorporation of this detection noise in our numerical simulations, we observe that the power and phase of the signal reconstructed from the noisy filtered heterodyne signal now deviate from their nominal numerical profiles, see Fig.~\ref{fig3_3}(a) and Fig.~\ref{fig3_3}(e). \\

By comparing Fig.\ref{fig3_3}(d) and Fig.\ref{fig3_3}(h) with Fig.~\ref{fig3_1}(d), we conclude that the detection noise of the PD contributes to the blurred appearance of the measured DOS. \\

In summary, the symmetric semi-circular shape of the DOS predicted by numerical simulations of an optical SG in the presence of damping is not perfectly replicated in the experiment, primarily due to noise and finite bandwidth effects. The asymmetric shape of the DOS measured in the experiment is primarily due to the finite bandwidth of the detection system. Detection noise contributes to the blurred appearance of the measured DOS. \\

\end{document}